\documentclass[10pt,amssymb,superscriptaddress,showkeys,twocolumn,prl]{revtex4-2}
\usepackage{graphicx}
\usepackage{dcolumn}
\usepackage{bm}
\usepackage{changes}
\usepackage{amsmath,amssymb}
\usepackage{titlesec}
\usepackage{lineno}
\usepackage{hyperref}

\makeatletter
\renewcommand{\@caption@fignum@sep}{\space} 
\makeatother
\titleformat{\section}[block]{\normalfont\bfseries\large}{\thesection}{1em}{}
\titlespacing*{\section}{0pt}{\baselineskip}{0.5\baselineskip}

\titleformat{\subsection}[block]{\normalfont\bfseries\small}{\thesubsection}{1em}{}
\titlespacing*{\subsection}{0pt}{\baselineskip}{0.5\baselineskip}

\hypersetup{
 colorlinks=true,
 linkcolor=blue,
 filecolor=blue,  
 urlcolor=blue,
 citecolor=blue,
}

\begin{document}

\title{Topological linear and nonlinear gap modes in defected dimer lattice with fourth-order diffraction}

\author{Xiaoyang Wang}
\affiliation{Department of Physics, Changzhi University, Changzhi, Shanxi 046011, China}

\author{Qidong Fu}
\affiliation{School of Physics and Astronomy, Shanghai Jiao Tong University, Shanghai 200240, China}

\author{Changming Huang}
\email{hcm123\_2004@126.com}
\affiliation{Department of Physics, Changzhi University, Changzhi, Shanxi 046011, China}

\date{\today}

\begin{abstract}
	In this study, we investigate the optical properties of linear and nonlinear topological in-phase, out-of-phase, and edge modes in a defected dimer lattice with fourth-order diffraction, encompassing their bifurcation characteristics, localized field distributions, and stability properties. Both focusing and defocusing nonlinearities are considered. Within the nontrivial regime, in-phase, out-of-phase, and edge modes can emerge within the gap. Numerical results reveal that three types of topological gap solitons bifurcate from their corresponding linear localized modes. The dimerization parameter can be tuned to drive the system from a trivial to a nontrivial configuration. We observe that as the strength of the fourth-order diffraction increases, the size of the spectral gap also enlarges, with this parameter effectively broadening the stability window for solitons. Our findings offer novel insights into the properties of both linear and nonlinear modes in topologically defected structures.
\end{abstract}

\maketitle

\section{Introduction}
\label{introduction}
The concept of photonic topological insulators, inspired by condensed matter physics~\cite{Hasan2010}, offers a novel platform for manipulating light waves with unique propagation characteristics, enabling unprecedented control over light routing and switching~\cite{Lu2014,Khanikaev2017}. 
By introducing magnetic fields~\cite{wang2008reflection, Wang2009, Fang2012} or constructing waveguide lattices with Floquet-engineered spiral waveguides~\cite{rechtsman2013photonic, Lumer2013, Leykam2016, Yang2020, ivanov2020edge, Ivanov2020, ivanov2021topological},
two-dimensional topological insulator structures can been employed and in which various linear and nonlinear unidirectional propagation phenomena have been observed~\cite{smirnova2020nonlinear, lin2023topological}.
Moreover, a novel class of higher-order topological insulators characterized by tunable waveguide spacing within and between unit cells has been designed, in which topological edge states, corner modes, and defect modes localized within the spectral gap have been extensively investigated both theoretically and experimentally~\cite{Noh2018, Li2019, Kirsch2021, Liu2021, Ren2023light, Ren2023apl, Huang2024, Kartashov2024}.
There are also one-dimensional ($1$D) topological systems that have edge states that localize at the ends of a lattice and their dynamics can be described by the Su-Schrieffer-Heeger (SSH) model~\cite{schomerus2013topologically, lang2014topologically, midya2018topological}.
One can know that these edge states in these $1$D systems are topologically protected, i.e., meaning they remain robust against various types of disorder.

Structural defects often introduce defect modes, whose existence and propagation characteristics have attracted considerable attentions~\cite{Makasyuk2006, Ye2008, Dong2010, Wang2020, Jovanovi2021, Sheng2022}. Small-scale and large-scale defects represent two common manifestations. 
In $1$D topological waveguide lattices, large-scale defects manifest mode properties similar to those of edge modes, primarily due to their distribution mirroring edge mode characteristics or emerging as a composite of two edge modes with negligibly weak coupling.
In small-scale defect lattices, the existence and stability of various defect solitons have been extensively investigated~\cite{yang2006defect, baugci2015fundamental, wang2011defect, lu2011defect, hu2012defect, zeng2021localized, chen2006surface, chen2010defect, zhu2010defect, lu2011surface, wang2024two, ivanov2021floquet, dong2010shaping}. These lattices encompass diverse configurations, including parity-time symmetric lattices with vacancy defects~\cite{yang2006defect, baugci2015fundamental, wang2011defect, lu2011defect, hu2012defect}, nonlinear lattice defects~\cite{zeng2021localized}, optical lattice surface defects~\cite{chen2006surface}, square/Kagome optical lattices with a defect~\cite{chen2010defect, zhu2010defect}, two-dimensional line defect lattices~\cite{lu2011surface, wang2024two, ivanov2021floquet}, and photonic lattices featuring flat-topped defects~\cite{dong2010shaping}. These studies have provided critical insights into the complex dynamics of defect-induced optical phenomena.

It is also worth noting that the optical properties of light beams within the optical lattices have been considered with second-order/fractional-order diffraction effects \cite{zhu2024two,chen2024dark,zeng2024solitons,zhu2025two}. However, in practical situations, higher-order diffraction/dispersion should also be considered~\cite{liu2009solitary, roy2009dispersive, tlidi2010high, kozyreff2009localized}.
In the regime of fourth-order diffraction/dispersion, diverse nonlinear phenomena emerge, encompassing intricate dynamics such as spectra of polarization-modulational instability~\cite{zambo2013impact}, Self-similar propagation of optical pulses~\cite{runge2020self}, localized dissipative structures~\cite{melchert2020dynamics}, and novel self-trapped soliton states~\cite{cole2014band, kruglov2020quartic, zanga2022generation, zhu2017gap, tiofack2018spatial, zhu2020gap, li2022dark, ge2014gap, turgut2024suppression}.

To the best of our knowledge, the properties of linear and nonlinear optical waves in $1$D topological waveguides with structural defects are not yet comprehensively understood, particularly when considering fourth-order diffraction effects. In this study, we investigate the optical properties of linear and nonlinear topological gap modes in a defected dimer lattice with fourth-order diffraction.
The dimerization parameter governs a structural transition between trivial and nontrivial configurations in this system. Within the nontrivial regime, in-phase and out-of-phase defect modes emerge alongside topological edge states residing in the gap. We have addressed the bifurcation characteristics, localized field distributions, and stability properties of three classes of gap solitons. Notably, the fourth-order diffraction term plays a dual role: it simultaneously modulates both the existence domains and stability regimes of these nonlinear localized states.

\section{Theoretical model}\label{sec2}
\begin{figure}
	\centering
	\includegraphics[width=0.9\columnwidth]{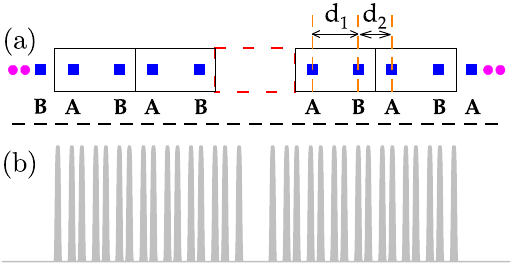}
	% \captionsetup{justification=centering}
	\caption{Panel (a) illustrates the scheme of the transverse waveguide array, featuring two waveguides labeled $\textbf{A}$ and $\textbf{B}$ within each unit cell. The centers of the waveguides are denoted by blue squares. The distance between the two waveguides within a unit cell is denoted as $d_1$, while the distance between corresponding waveguides in adjacent unit cells is labeled as $d_2$. A unit cell defect is present at the center of the waveguide array, indicated by a red dashed box. Panel (b) displays the complete waveguide array.} 
	\label{fig1}  
\end{figure}

We consider the propagation of a light beam along the $z$-axis in a 1D finite optical lattice, and the dynamics can be described by the following dimensionless nonlinear Schr\"{o}dinger equation~\cite{ge2014gap,turgut2024suppression}:
\begin{equation}\label{eq1}
	i\frac{\partial q}{\partial z} +\frac{1}{2}\frac{\partial^2 q}{\partial x^2}-\chi   \frac{\partial^4 q}{\partial x^4}+V(x)q+\sigma \left |  q\right |^2q=0,  
\end{equation}
Here, $q(x,z)$ represents the normalized wave packet of the electric field, with $x$ and $z$ denoting the transverse coordinate and propagation direction, respectively. The parameter $\chi>0$ quantifies the strength of the fourth-order diffraction, with the conventional Schr\"{o}dinger equation emerging when $\chi=0$. $\sigma$ denotes the coefficient of the cubic nonlinearity, for focusing nonlinearity, $\sigma=+1$, whereas for defocusing nonlinearity, $\sigma=-1$.
The function $V(x)$ describes the 1D dimer lattice, and is expressed as $V(x)=p\sum_{m=1}^{m=n} \exp \left [ -\left ( x-x_m \right )^6/d^6  \right ] $, where $p$ represents the lattice depth, $d$ is the width of a single waveguide, and $x_m$ denotes the center of the dimer lattice. Within each dimer unit cell, there are waveguides $A$ and $B$. The distance between these two waveguides within a unit cell is denoted as $d_\text{AB} = d_1$, while the distance between waveguides in neighboring unit cells is denoted as $d_\text{BA} = d_2$. 
The center of the waveguides, $x_m$, can be determined by the ratio $\gamma = d_1 / d_2$ and the length of a single unit cell $L = d_1 + d_2$. Specifically, the dimer lattice we constructed consists of $17$ unit cells (i.e., $n = 34$), with the center of the unit cell array located at $x = 0$. Without loss of generality, a unit cell defect was introduced at the center of the lattice. The proposed dimer lattice structure illustrated in Fig.~\ref{fig1} supports a range of linear modes, including topological edge modes and defect modes.

In subsequent discussions, we set the parameters as $p = 4.0$, $d = 0.5$, and $L=4.7$, while exploring the properties of linear and nonlinear modes by varying $\chi$, $\gamma$, and $\sigma$.

\begin{figure}
	\centering
	\includegraphics[width=0.9\columnwidth]{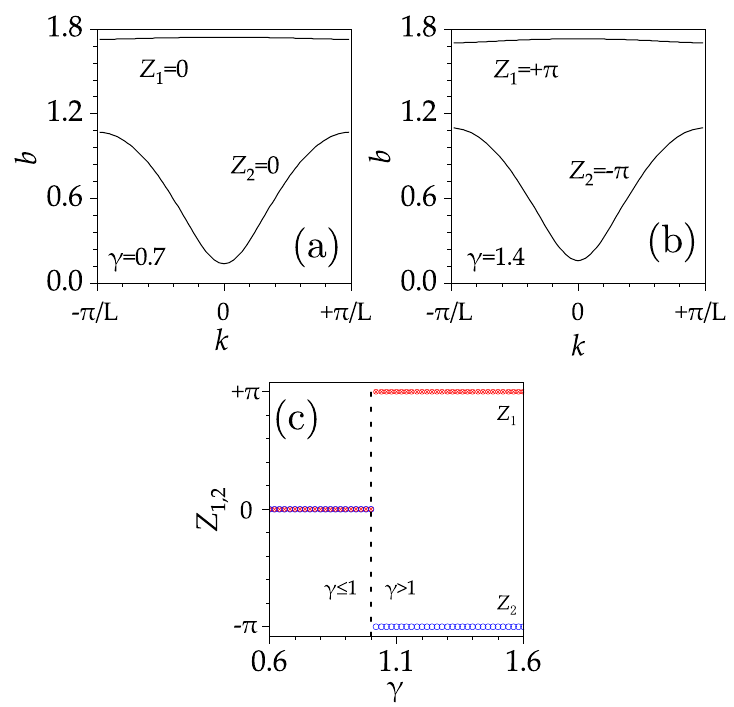}
	% \captionsetup{justification=centering}
	\caption{At $\chi=0.5$, the Floquet-Bloch spectrum $b(k)$ is shown for $\gamma=0.7$ (a) and $\gamma=1.4$ (b). Panel (c) displays the Zak phase $Z_{1,2}$ for the first and second bands as functions of $\gamma$, represented by red crossed circles and blue circles, respectively. }
	\label{fig2}  
\end{figure}

\begin{figure}%[!t] %
	\centering
	\includegraphics[width=0.95\columnwidth]{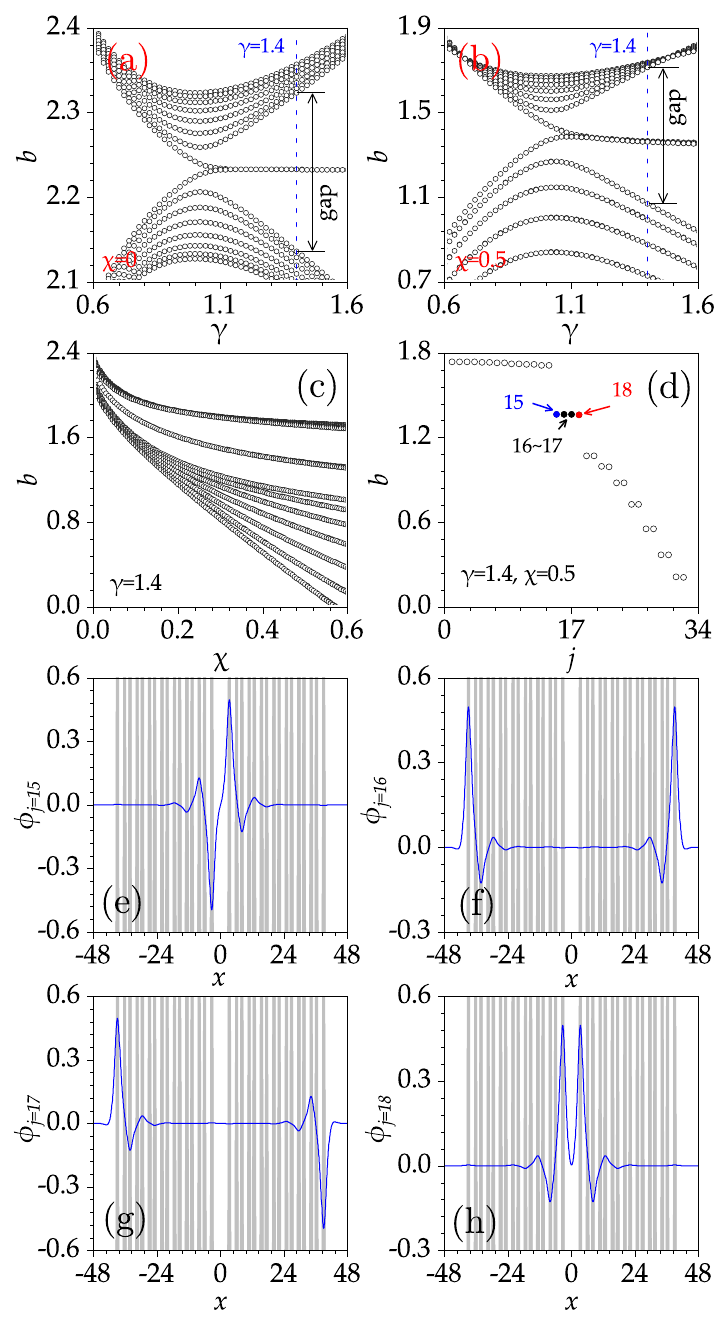}
	\caption{The eigenvalues for various ratios of $\gamma$ at $\chi = 0$ (a) and $\chi=0.5$ (b), and different values of $\chi$ at $\gamma = 1.4$ (c). In (a) and (b), blue dashed lines indicate $\gamma = 1.4$, with the gap size marked by double-headed arrows. Panel (d) shows the eigenvalues at $\gamma = 1.4$ and $\chi = 0.5$. Panels (e), (f), (g) and (h) depict the linear out-of-phase, edge, and in-phase modes associated with the solid-marked points in panel (d). The embedded gray patterns in (e), (f), (g) and (h) represent the waveguide arrays. 
	} 
	\label{fig3}  
\end{figure}

\begin{figure}[htp] %
	\centering
	\includegraphics[width=0.95\columnwidth]{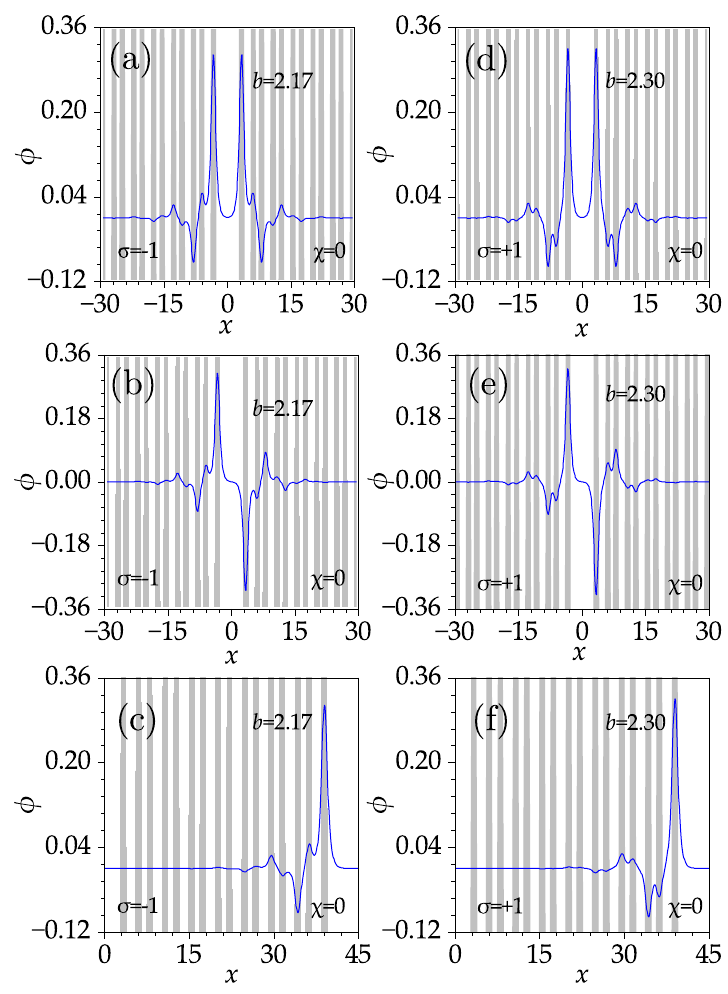}
	% \captionsetup{justification=centering}
	\caption{Profiles of in-phase (a, d), out-of-phase (b, e), and edge mode solitons (c, f) are plotted for both defocusing (a-c) and focusing (d-f) nonlinearities. $b=2.17$ in panels (a-c) and $b=2.30$ in panels (d-f).
		Parameters $\chi = 0$ and $\gamma=1.4$ in all panels.
	} 
	\label{fig4}  
\end{figure}

\begin{figure}[ht] %
	\centering
	\includegraphics[width=0.95\columnwidth]{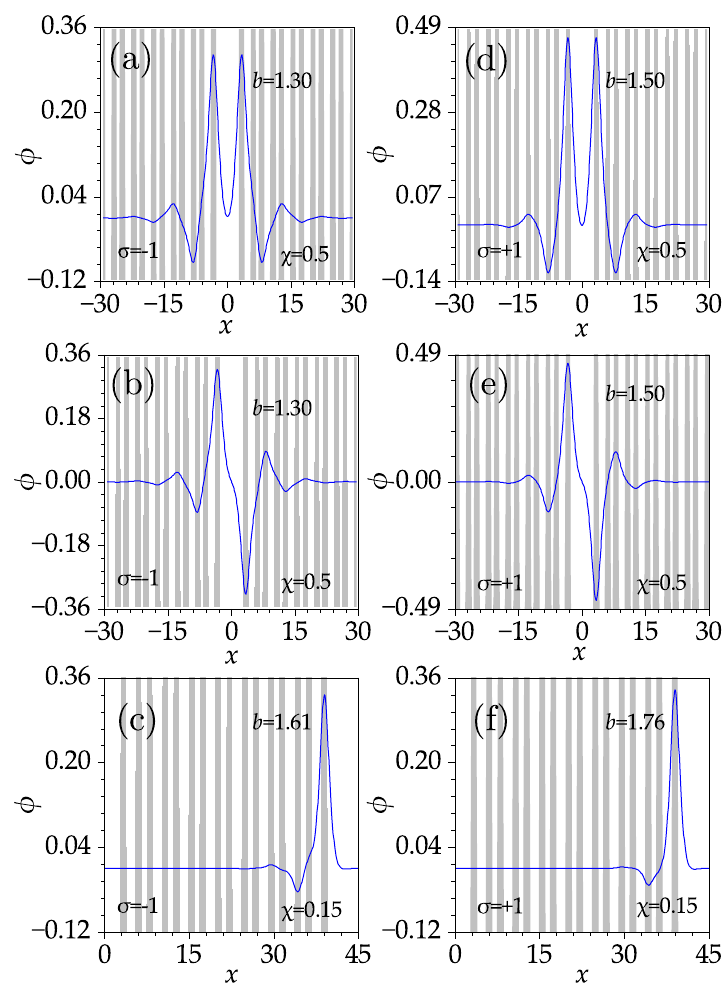}
	% \captionsetup{justification=centering}
	\caption{Profiles of in-phase (a, d), out-of-phase (b, e), and edge mode solitons (c, f) are plotted for both defocusing (a-c) and focusing (d-f) nonlinearities. $b=1.3$ and $\chi = 0.5$ in panels (a, b) and $b=1.5$ and $\chi = 0.5$ in panels (d, e), $b=1.61$ and $\chi = 0.15$ in panel (c) and $b=1.76$ and $\chi = 0.15$ in panel (f). $\gamma=1.4$ in all panels.
	} 
	\label{fig5}  
\end{figure}

\section{Results and discussion}\label{sec3}
In the proposed lattice structure, the dimerization parameter governs the ratio of intra- and inter-cell waveguide distances, offering a versatile mechanism to modulate the topological phase of optical lattices. Revealing the intricate interplay between topological phases and localized modes is crucial for advancing our fundamental understanding of these complex systems.
We calculate the global topological invariant of the system, specifically the Zak phase~\cite{vanderbilt2018berry}, which is defined as
${Z_{n}=\oint_{\mathrm{BZ}}\left\langle \phi_{n k} \mid i \partial_{k} \phi_{n k}\right\rangle}$,
where $n$ denotes the band index, $k$ represents the Bloch wave number, the notation $\oint_{\mathrm{BZ}}$ indicates an integral taken around the loop formed by the $1$D Brillouin zone, and $\phi_{n k}$ represents the complex periodic function.

The Bloch wave takes the form $q(x,z)=\phi_k(x)e^{ibz+ikx}$, with $b$ as the linear eigenvalue, and $\phi_k(x)=\phi_k(x+L)$ represents a complex periodic function. By substituting this expression into linear version of Eq.~(\ref{eq1}), we obtain
\begin{equation}
	\begin{split}
		&b \phi_k=+\frac{1}{2}\left(\frac{\partial^{2} \phi_k}{\partial x^{2}}+2 i k \frac{\partial \phi_k}{\partial x}-k^{2} \phi_k\right)+V(x)\phi_k\\
		&-\chi\left(\frac{\partial^{4} \phi_k}{\partial x^{4}}+4 i k \frac{\partial^{3} \phi_k}{\partial x^{3}}-6 k^{2} \frac{\partial^{2} \phi_k}{\partial x^{2}}-4 i k^{3} \frac{\partial \phi_k}{\partial x}+k^{4} \phi_k\right).
	\end{split} \nonumber
\end{equation}
The Floquet-Bloch spectrum $b(k)$ can be solved numerically. By collecting all $\phi_k(x)$ and inserting them into the expression of the Zak phase, the topological invariant of the system can be obtained.

Representative Floquet-Bloch spectra are shown in Figs. \ref{fig2}(a) and \ref{fig2}(b). Notably, for infinite periodic lattices, the $b(k)$ curves exhibit minimal variation for $\gamma=0.7$ and $\gamma=1.4$. However, for cases with $\gamma<1$ and $\gamma>1$, the boundary waveguide distributions in truncated finite waveguides exhibit substantial differences, leading to distinct Zak phases [Fig. \ref{fig2}(c)].

Elucidating the fundamental properties of linear modes is essential for comprehending nonlinear behavior. We examined the eigenvalue spectrum of Eq. (\ref{eq1}) by excluding the nonlinear term (i.e., setting $\sigma = 0$). 
The eigenvalues and modal distributions of the linear modes are depicted in Fig.~\ref{fig3}.

The eigen-spectrum corresponding to the ratio $\gamma$ of the distances between waveguides within a unit cell and those between adjacent unit cells are shown in Figs.~\ref{fig3}(a) and~\ref{fig3}(b). We find that when $\gamma \neq 1$, the structure exhibits a large spectral gap. As the parameter $\gamma$ deviates from $\gamma=1$, the spectral gap gradually increases in width. Moreover, for a given $\gamma$ (e.g., $\gamma=1.4$), the corresponding spectral gap increases progressively with increasing $\chi$, as quantitatively illustrated in Fig.~\ref{fig3}(c).
For $\gamma > 1$, there are four isolated eigenvalues within the gap (as detailed in Fig.~\ref{fig3}(d)). Notably, the modes associated with these four eigenvalues manifest as linearly localized modes, including out-of-phase defect mode (with index $j=15$), two topological edge modes (with indices $j=16$ and $j=17$), and in-phase defect mode (with indices $j=18$) [Figs.~\ref{fig3}(e), (f), (g), and (f)].

\begin{figure}%[htbp] %
	\centering
	\includegraphics[width=0.95\columnwidth]{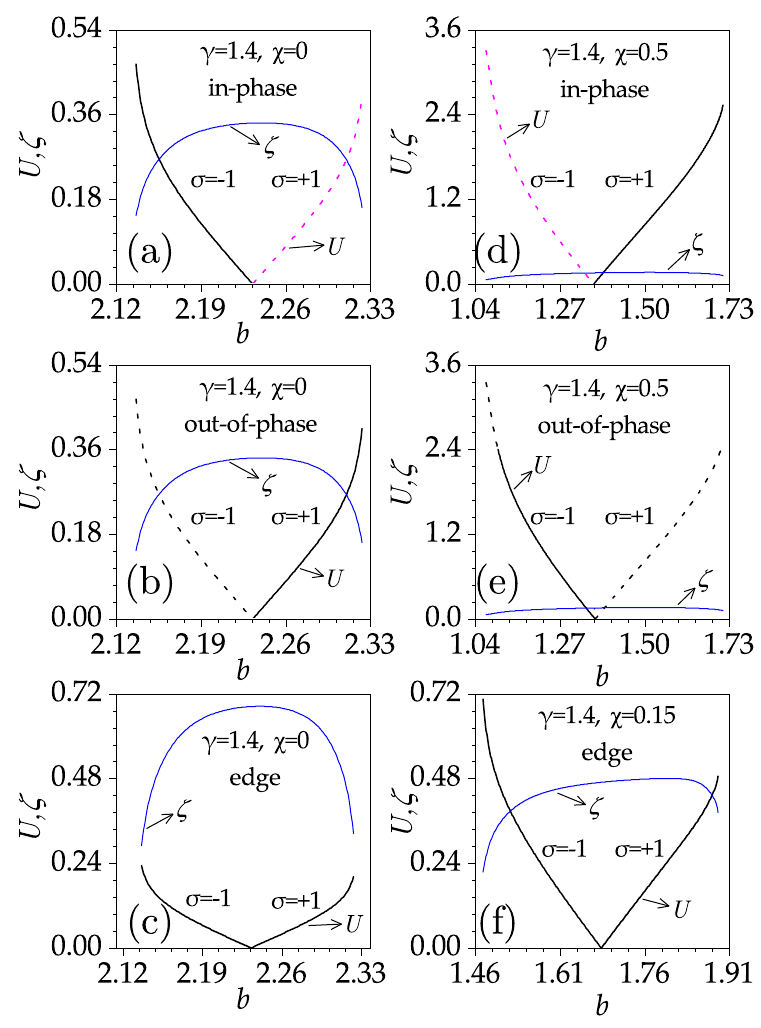}
	% \captionsetup{justification=centering}
	\caption{Power $U$ and form factor $\zeta$ of the soliton families bifurcating from the linear in-phase (a, d), out-of-phase (b, e), edge modes (c, f).
		In (a) and (b), magenta lines characterize solitons as remarkably robust, exhibiting minor peak fluctuations (weak instability regions). In (c-f), solid lines denote stable evolution regions, while dashed lines represent unstable regions. $\chi=0$ in panels (a-c), $\chi=0.5$ in panels (d, e), and $\chi=0.15$ in panel (f).
	} 
	\label{fig6}  
\end{figure}

\begin{figure}[b]%[htbp] %
	\centering
	\includegraphics[width=0.95\columnwidth]{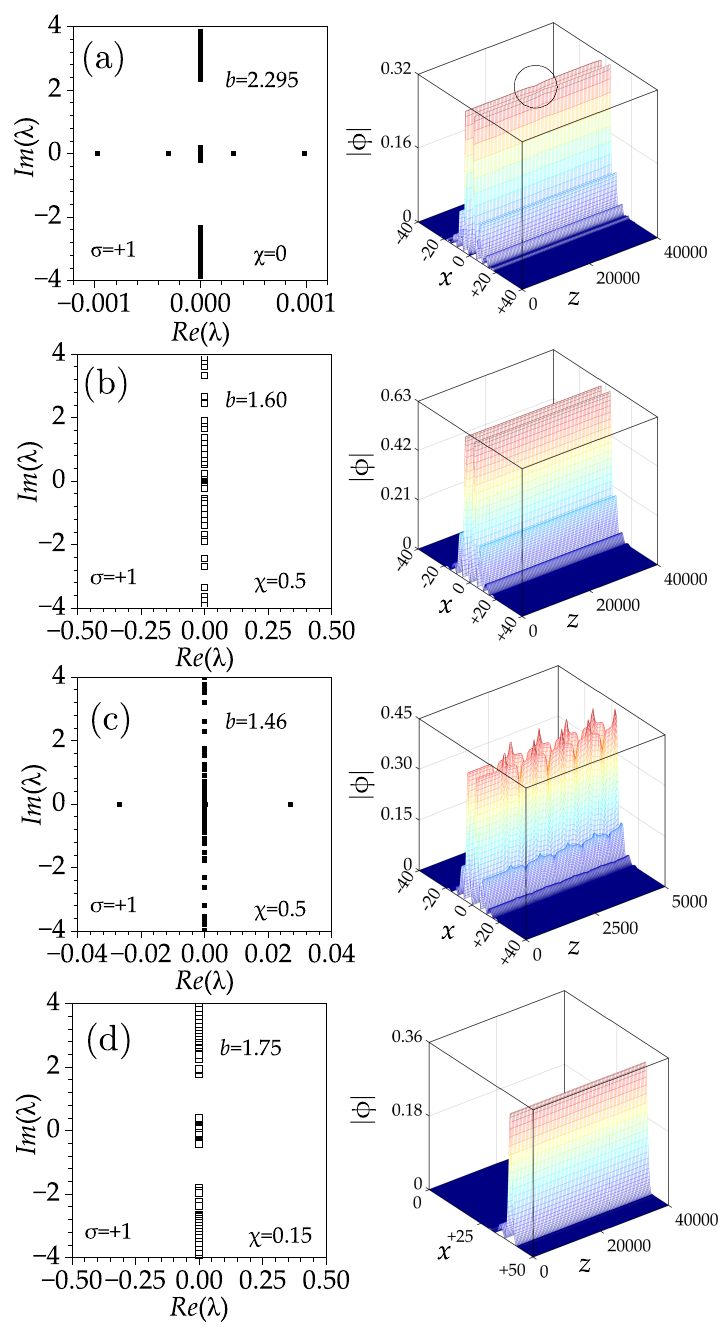}
	\caption{Representative linear stability spectra (left) and their corresponding propagation evolution patterns (right). (a) $\chi=0,b=2.295$ for weakly unstable in-phase soliton; (b) $\chi=0.5, b=1.60$ for stable in-phase soliton; (c) $\chi=0.5, b=1.46$ for unstable out-of-phase soliton; (d) $\chi=0.15, b=1.75$ for stable edge soliton. The minor peak fluctuations are marked by the circle in the right panel of (a).
	} \label{fig7}  
\end{figure}

Next, we investigate the properties of the soliton families that bifurcate from the defect mode and edge mode. The soliton solution is of the form $q(x, z) = \phi(x)\exp(ibz)$, where $\phi(x)$ represents the profile of the soliton and $b$ is the propagation constant. By substituting this expression into Eq.(\ref{eq1}), we obtain the ensuing nonlinear equation: 
\begin{equation}\label{eq2}
	\frac{1}{2}\frac{\partial^2 \phi}{\partial x^2}-\chi   \frac{\partial^4 \phi}{\partial x^4}+V(x)\phi-b\phi+\sigma \left |  \phi\right |^2\phi=0,  
\end{equation}
One can solve this equation by using the Newton iteration method.

Firstly, we analyze the distribution characteristics of in-phase, out-of-phase, and edge mode solitons under the conventional Schr\"{o}dinger equation (i.e., $\chi=0$), as typically depicted in the profile of Fig.~\ref{fig4}.
We find that the maximum amplitudes of the in-phase [panels (a) and (d) of Fig.~\ref{fig4}] and out-of-phase [panels (b) and (e) of Fig.~\ref{fig4}] defect solitons are situated on the waveguides adjacent to the defect cell. In contrast, edge solitons [panels (c) and (f) of Fig.~\ref{fig4}] are predominantly localized at the waveguide near the boundary.
In this regime, defect solitons exhibit corresponding peaks in several waveguides near the center. Specifically, within the region $x>0$ (or $x<0$), under defocusing nonlinearity, the peaks supported by the two waveguides within a single unit cell are in-phase, while the peaks supported by adjacent unit cells are out-of-phase. Under focusing nonlinearity, the peaks supported by the two waveguides within a single unit cell are out-of-phase, while the peaks supported by adjacent unit cells remain out-of-phase. Therefore, the distribution of soliton peaks is closely associated with the characteristics of nonlinear signals.

Furthermore, when the fourth-order diffraction effects are considered, the profile distributions of three types of soliton families exhibit differences compared to the case when $\chi=0$ [Fig.~\ref{fig5}]. Specifically, at $\chi=0.5$, we find that peaks do not localize in all waveguides, but only one waveguide within a unit cell supports the peak, with out-of-phase peaks existing between unit cells. Under both types of nonlinearity, the soliton profiles are similar but also exhibit differences, with the distinction manifesting in the sharp transition of the secondary maximum between defocusing and focusing nonlinearities.

Three types of soliton families parameterized by propagation constant $b$ are depicted in Fig.~\ref{fig6}.
For defocusing nonlinearity ($\sigma=-1$), the propagation constant of solitons bifurcating from linear modes is situated between the lower boundary of the spectral gap and the eigenvalue of the corresponding linear mode. In contrast, for focusing nonlinearity ($\sigma=+1$), the propagation constant ranges from the eigenvalue of the linear mode to the upper boundary of the spectral gap [Fig.~\ref{fig6}].
We can also find that the soliton's power [defined as $U=\int_{-\infty}^{+\infty}|\phi(x)|^2dx$] diminishes as the parameter $b$ increases, 
asymptotically approaching zero near the critical value ($b_\mathrm{lin}=2.233$ at $\chi=0$, and $b_\mathrm{lin}=1.36$ at $\chi=0.5$) for $\sigma=-1$. In contrast, for $\sigma = +1$, the soliton's power exhibits a monotonic increase with respect to its propagation constant.

Figure~\ref{fig6} also quantitatively characterizes the form factor of the soliton family [see the blue lines, defined as $\zeta=U^{-2}\int_{-\infty}^{+\infty}|\phi(x)|^4dx$], which represents the localization degree of the soliton. The larger the form factor, the more localized the soliton is. Clearly, for $\sigma = -1$, the soliton’s form factor increases, indicating a progressive narrowing of the soliton profile. For $\sigma = +1$, the soliton’s form factor exhibits an initial increase, reaches a maximum, and then decreases, indicating a complex evolution of its spatial profile with increasing power.
Notably, as the soliton becomes more localized within an individual waveguide, its oscillatory tail starts to extend into multiple unit cells near the upper band edge, resulting in a broader characteristic.

It should be noted that, regardless of focusing or defocusing nonlinearity, as the propagation constant $b$ nears $b_\text{lin}$, the power of the solitons tends to zero, and the form factor exhibits a smooth transition from the $\sigma=-1$ to $\sigma=+1$ regions. These characteristics confirm that the three types of soliton families we obtained are power thresholdless and bifurcate from their corresponding linear modes.

The stability of topologically gap solitons is a key concern we are investigating. To analyze the stability of three types of soliton families, we perform a detailed stability analysis by introducing perturbations to the static soliton solution. The perturbed solution takes the form: $q(x,z)=[\phi(x)+u(x)\exp(\lambda z)+iv(x)\exp(\lambda z)]\exp(ibz)$, where $\lambda$ is a complex rate, and $u$ and $v$ denote the real and imaginary components of the perturbation, respectively. These components may evolve as the propagation distance increases.
By substituting this perturbation form into Eq. (\ref{eq1}), we obtain the following linearly coupled equations: \begin{equation}\label{eq3}
	% \left\{
	\begin{aligned}
		&\lambda u=-\frac{1}{2} \frac{\partial^2{v}}{\partial{x^2}}+\chi\frac{\partial^4{v}}{\partial{x^4}}+bv-Vv-\sigma\phi^2v, \\
		&\lambda v=+\frac{1}{2} \frac{\partial^2{u}}{\partial{x^2}}-\chi\frac{\partial^4{u}}{\partial{x^4}}-bu+Vu+3\sigma\phi^2u.
	\end{aligned}
	% \right.
\end{equation}
The above coupled equations can be numerically solved using the difference method combined with an eigenvalue solver. When $\mathcal{M}ax\{\mathcal{R}e(\lambda)\}>0$, the soliton is unstable, and vice versa.

Linear stability analysis and direct propagation simulations reveal that, for $\chi = 0$, in-phase defect gap solitons are stable within their existence region for $\sigma = -1$, and remarkably robust with only minor peak fluctuations for $\sigma = +1$ [Fig.~\ref{fig6}(a)]. 
Out-of-phase solitons exhibit opposite stability properties compared to in-phase counterparts under two types of nonlinearities [Fig.~\ref{fig6}(b)].
When $\chi = 0.5$, the in-phase defect solitons are unstable for $\sigma = -1$, but stabilize for $\sigma = +1$ within their existence region [Fig.~\ref{fig6}(d)], while the out-of-phase defect solitons can be stable for $\sigma = -1$, but unstable for $\sigma =+1$ [Fig.~\ref{fig6}(e)].
The parameter $\chi$ plays a crucial role in modulating the stability region of gap solitons [compared to Figs.~\ref{fig6}(a), \ref{fig6}(b), \ref{fig6}(d), and \ref{fig6}(e)]. In general, for $\sigma = +1$, larger $\chi$ favors the stability of in-phase gap solitons, while smaller $\chi$ promotes stable out-of-phase gap solitons. Conversely, for $\sigma = -1$, smaller $\chi$ stabilizes in-phase solitons, whereas larger $\chi$ favors stable out-of-phase solitons.
Moreover, edge solitons are stable for small $\chi$ in their existence domain [Figs.~\ref{fig6}(c) and \ref{fig6}(f)].

Typical spectrum and propagation patterns are shown in Fig.~\ref{fig7}. At $\chi= 0$, in-phase and out-of-phase solitons with a maximum $\mathcal{R}e(\lambda)$ of $\sim 10^{-3}$ exhibit small fluctuations after propagating over a considerable distance [Fig.~\ref{fig7}(a)].
When $\chi= 0.5$, the spectrum of a stable in-phase soliton exhibits all real parts equal to zero, enabling its propagation distance up to $z = 4\times 10^{4}$ without any distortion [Fig.~\ref{fig7}(b)]. 
Figure~\ref{fig7}(c) depicts a representative case of an unstable out-of-phase soliton, where the field distribution undergoes rapid oscillations across adjacent lattice channels after a short propagation distance. The pronounced instability is characterized by power transfer to neighboring lattice sites, thereby disrupting the soliton's intrinsic spatial field configuration. The evolutions of topological edge states can also be anticipated, with field confinement strictly localized to the boundary waveguide [Fig.~\ref{fig7}(d)].

\section{Conclusion}\label{sec4}

In the one-dimensional defected dimer lattice, we have discerned the presence of localized modes within the gap, encompassing in-phase, out-of-phase, and edge modes. These modes give rise to corresponding solitons that bifurcate from their linear counterparts under both focusing and defocusing nonlinear regimes. Within the framework of the conventional Schr\"{o}dinger equation, in-phase solitons demonstrate complete stability for defocusing nonlinearity, whereas out-of-phase solitons exhibit complete stability in their entire domain of existence under focusing nonlinearity. Conversely, in-phase solitons under focusing nonlinearity and out-of-phase solitons under defocusing nonlinearity display considerable robustness in their existence region, with only minor peak amplitude variations observed during propagation. Topological edge solitons, on the other hand, are found to be stable within the gap.
When fourth-order diffraction effects are present, the field distribution of gap solitons differs significantly from cases where such effects are absent. Specifically, the peaks do not localize in all waveguides; instead, only one waveguide within a unit cell supports the peak, with out-of-phase peaks existing between unit cells. 
Importantly, fourth-order diffraction effects can effectively modulate the size of the spectral gap and the stability region of solitons. 
Our results significantly advance the comprehension of linear and nonlinear mode dynamics within topological defected optical lattices.

\section*{Acknowledgements}
This work was supported by the Applied Basic Research Program of Shanxi Province (202303021211191), the National Natural Science Foundation of China (No. 12404385) and China Postdoctoral Science Foundation (No. BX20230218, No. 2024M751950).

\end{document}